\documentclass[sigconf]{acmart}

\usepackage{booktabs} 
\usepackage[mode=buildnew]{standalone}
\usepackage{xcolor}

\usepackage{hyperref}

\usepackage{tikz}
\usetikzlibrary{calc}

\usepackage{standalone}

\usepackage{listings}

\usetikzlibrary{shadows}
\usepackage{pgfplots}
\pgfplotsset{compat=1.14}

\usepackage{multirow}

\usepackage[textsize=tiny]{todonotes} 
\newcommand{\AbhishekOld}[1]{}

\setcopyright{acmlicensed}

\acmDOI{}

\acmISBN{}

\acmConference[M4IoT]{4th Workshop on Middleware and Applications for the Internet of Things}{December 2017}{Las Vegas, USA} 
\acmYear{2017}
\copyrightyear{2017}

\acmPrice{15.00}

\begin{document}
\setlength{\marginparwidth}{1.5cm}

 \title[PlaTIBART: a Platform for Transactive IoT Blockchain Applications with Repeatable Testing]{PlaTIBART: a Platform for Transactive IoT Blockchain Applications with Repeatable Testing}


\author{Michael A. Walker, Abhishek Dubey, Aron Laszka, and Douglas C. Schmidt}
\affiliation{\institution{Vanderbilt University}}
\email{{michael.a.walker.1, abhishek.dubey, a.laszka, d.schmidt}@vanderbilt.edu, }

\renewcommand{\shortauthors}{M. Walker et al.}

\begin{abstract}
{\em
With the advent of blockchain-enabled IoT applications, there is an increased need for related software patterns, middleware concepts, and testing practices to ensure adequate quality and productivity. IoT and blockchain each provide different design goals, concepts, and practices that must be integrated, including the distributed actor model and fault tolerance from IoT and transactive information integrity over untrustworthy sources from blockchain.
\AbhishekOld{for example, .. are we referring to distributed actor model and transactive information integrity over untrustworthy sources. It will be good to be precise}
Both IoT and blockchain are emerging technologies and both lack codified patterns and practices for development of applications when combined.
\AbhishekOld{This is first time we have used the world cyber-physical here}
This paper describes PlaTIBART, which is a platform for transactive IoT blockchain applications with repeatable testing that combines the Actor pattern (which is a commonly used model of computation in IoT\AbhishekOld{How does this relate to IoT, you can say a commonly used model of computation in IoT}) together with a custom Domain Specific Language (DSL) and test network management tools. We show how PlaTIBART has been applied to develop, test, and analyze fault-tolerant \AbhishekOld{Fault-tolerant is being described very late. you should describe this earlier as an example of the well-defined patterns} IoT blockchain applications.}

\end{abstract}

%
%




\maketitle

\section{Introduction}

Interest in---and commercial adoption of---blockchain technology has increased in recent years~\cite{TheTruthAdoption:online}. For example, blockchain adoption in the financial industry has yielded market capitalization surpassing \$75 billion USD~\cite{BitcoinPrices:online} for Bitcoin and \$36 billion USD for Ethereum ~\cite{EthereumPrices:online}.  One reason for this growth stems from blockchains' combination of existing technologies to enable  the interoperation of non-trusted parties in a decentralized, cryptographically secure, and immutable ecosystem without the need of a trusted central authority.

During roughly the same time, the increased proliferation of IoT devices has motivated the need for transactional integrity due to the transition of IoT devices from just being smart-sensors to being active participants that impact their environment via communication, decision making, and physical actuation. These abilities require transactional integrity to provide auditing of actions made by potentially untrusted networked 3rd party IoT devices. The demand for transactional integrity in IoT devices that simultaneously leverage  blockchain features (such as decentralization, cryptographic security, and immutability) has motivated research on creating transactive IoT blockchain applications~\cite{Bogner2016,Buccafurri2017}.

Blockchain deployments (and specifically Ethereum, which is the focus of this paper due to its large installed base, and its powerful smart contract language) are generally managed via programs that have different modes in which they can operate. They broadly fall into Command-Line Interfaces (CLI), RPC APIs, or creating Graphical Interfaces via the use of HTML pages and JavaScript code~\cite{EthereumAPInterfaces:online}. These interfaces provide standard means to either run Ethereum applications within the clients themselves, or to interface other applications with the Ethereum clients. 

In practice, however, the existing blockchain deployment interfaces lack built-in fault tolerance, most notably for either network communication errors or application execution faults. Moreover, Ethereum clients are deployed manually since no official manager exists for them. As a result, developers can---and do~\cite{ethereumLostCoinAPI:online}---lose all of their Ether (Ethereum's digital currency) due to insecure client configurations. This problem is compounded by the fact that Ethereum's clients do not warn of this risk within their built-in help feature, and instead rely upon online documentation to warn developers. 

Addressing this problem requires patterns and tools that enable the deployment of blockchain clients in a repeatable and systematic way.
This requirement becomes even more important when integrating IoT blockchain applications (ITBAs). The IoT component of ITBAs add other requirements atop traditional blockchain applications due to their interactions with the physical environment and increased privacy concerns, \emph{e.g.}, thus preventing leakage of personal data, such as  energy usage that would reveal a user's activity patterns in their home~\cite{gubbi2013IoT}.  

Moreover, ITBAs may not only communicate over the blockchain, but may also use off-blockchain communications via TCP/IP or other networking protocols for the following reasons:
\begin{itemize}
\item  There are interactions with the physical environment that might require communication with sensors and/or actuators.  For example, a user's smart-meter might communicate wirelessly with their smart-car's battery to activate charging based on current energy production/cost considerations. 
\item The distributed ledger (which makes an immutable record of transactions in blockchain) is public, so it is common to only  include information within transactions that can safely be stored publicly. In particular, if some or all data from a transaction must be kept secret for privacy or any other reasons the transaction can, instead, contain the meta-data and a cryptographic hash of the secret data.  Private information must, therefore, be communicated off-blockchain while still preserving integrity by storing meta-data and hash information on the blockchain ledger.
\item Management tasks such as: updates, monitoring, calibration, debugging, or auditing may require off-blockchain communication (with possible on-blockchain components for logging). Currently, these management tasks are done manually in conventional blockchain ecosystems. Similar to the need for a systematic means of deploying apps in a blockchain network, there is a need to systematically configure the network topology between all components of ITBAs.
\end{itemize}









This paper presents the structure and functionality of PlaTIBART, which is a \textit{Platform for Transactive IoT Blockchain Applications with Repeatable Testing} that provides a set of tools and techniques for enhancing the development, deployment, execution, management, and testing of ITBAs. In particular, we describe a pattern for developing ITBAs, a Domain Specific Language (DSL) for defining a private blockchain deployment network, Actor components upon which the application can be deployed and tested, a tool using these DSL models to manage deployment networks in a reproducible test environment, and interfaces that provide fault tolerance via an application of the \textit{Observer} pattern. 

The remainder of this paper is organized as follows: 
Section \ref{system_model} explains the system model underlying PlaTIBART and describes the scenario of transactive energy used in this paper to motivate the need for ITBAs;
Section \ref{stateOfArt} reviews the current state-of-the-art regarding IoT and blockchain integration;
Section \ref{proposedArch} illustrates our proposed ITBA architecture and shows how we use the \textit{Actor} pattern to construct our solution, the DSL we created, and the network manager script we created to generate  test networks for ITBAs; Section \ref{empirical} examines our experimental testbed configuration and analyzes our results; Section \ref{lessonslearned} summarizes lessons learned, while implementing our proposed architecture; and Section \ref{conclusion} presents concluding remarks and outlines future work. 

\section{System Model}
\label{system_model}

This section explains the system model underlying PlaTIBART and describes the use case scenario of transactive energy we employ in this paper to motivate the need for ITBAs.
%
Based on our experience developing decentralized apps (DApps) for blockchain ecosystems~\cite{Zhang2016,dubey2017resilience},  three key capabilities are essential for DApps to function effectively in an ITBA ecosystem: traditional IoT computations and interactions should be supported, information should be robustly sorted in a distributed database, and a system-wide accepted sequential log of events should be provided. Each requirement can be delegated to a separate layer in a three-tiered architecture.  The first tier is the IoT middleware layer that facilitates communication between networked devices, which can be addressed by existing IoT middleware, such as RIAPS~\cite{RIAPS_IEEE_ISORC_2017}. The second tier is a distributed database layer. The third tier is a sequential log of events layer, which can be solved by blockchain integration. 

PlaTIBART provides an architecture for coordinating all these layers in a fault tolerant manner, along with tools for repeatable testing at scale. It leverages the Actor model~\cite{lee2003actor} to integrate these three layers. Each layer is composed of components that accomplish their designated layer-dependent tasks. These components are then combined into a single actor that can interact with each layer and other actors in the network, as described in Section~\ref{proposedArch}.

\subsection{Case Study: Transactive Energy System}
\label{case_study}


Transactive Energy Systems (TES) have emerged in response to the shift in the power industry away from centralized, monolithic business models characterized by bulk generation and one-way delivery toward a decentralized model in which end users play a more active role in both production and consumption \cite{NIST_TE,Gridwise}.
The GridWise Architecture Council defines TES as ``a system of economic and control mechanisms that allows the dynamic balance of supply and demand across the entire electrical infrastructure, using value as a key operational parameter'' \cite{Gridwise}. 

In this paper, we consider a class of TES that operates in a grid-connected mode, meaning the local electric network is connected to a Distribution System Operator (DSO) that provides electricity when the demand is greater than what the local-network can generate. The main actors are the consumers, which are comprised primarily of residential loads, and prosumers who operate distributed energy resources, such as rooftop solar batteries or flexible loads capable of demand/response. Additionally, the DSO manages the grid connection of the network. Such installations are equipped with an advanced metering infrastructure consisting of TES-enabled smart meters.
Examples of such installations include the Brooklyn Microgrid Project \cite{BrooklynMicrogrid} and the Sterling Ranch learning community~\cite{SterlingRanch}. A key component of TES is a transaction management platform (TMP), which handles market clearing functions in a way that balances supply and demand in a local market.


\section{Analysis of State-of-the-Art}
\label{stateOfArt}

This section reviews the state-of-the-art in IoT and blockchain integration,  focusing on testing. Prior work~\cite{christidis2016blockchains} has shown that IoT and blockchain can be integrated, allowing peers to interact in a trustless, auditable manner via the use of blockchain as a resilient, decentralized, and peer-to-peer ledger. Work has also been done on the topics of security and privacy of IoT and Blockchain integrations~\cite{dorri2017blockchain,ouaddah2017towards}. Beyond that, work has focused on formal verification of smart contracts~\cite{kumaresan2014use}, and how to write smart contracts ``defensively''~\cite{delmolino2016step} to avoid exceptions when multiple contracts interact. The current state-of-the-art with respect to testing, however, is lacking because blockchains are infrequently tested at scale in a systematic and repeatable manner, so we focus on that below.

\subsection{Testing IoT Blockchain Systems}


Popular blockchain ecosystems, such as Bitcoin and Ethereum, suffer from design limitations that prevent their direct application to IoT. In particular, transaction-confirmation times are relatively long (around tens of minutes) and variable on public blockchain networks due largely to their proof-of-work algorithms. Likewise, IoT devices have limited processing power and storage capabilities, which must be accounted for and tested~\cite{IoTandBl20:online} to ensure constraints are met.

Prior work on testing of IoT blockchain systems generally fall into two categories: (1) their test implementation has a single client and one or more smart contracts or (2) they focus purely on theoretical aspects and discuss future work implementing a test example. For example, Beck et.\ al~\cite{beck2016blockchain} discusses their implementation, but apparently (since it is not discussed in detail) the implementation only uses a single client, two smart contracts, and no additional transactions on the ledger. Conversely, Simic et.\ al \cite{simicCaseStudyIoTBlockchainHealthcare2017}  presents a purely theoretical paper where they discuss IoT and blockchain powered healthcare at a high level, without addressing privacy or any of the many other significant implementation difficulties. 

\subsection{Testing Repeatability}

The importance of integration and regression testing in software development has been well-established for over 20 years~\cite{rothermel2001prioritizing,agrawal1993incremental,leung1990study}. Integration and regression testing of distributed systems has been improved via network emulation testbeds, such as DETERLAB~\cite{mirkovic2012teachingDETERLAB} and Emulab~\cite{siaterlis2013useEmulab}. These testbeds provide mechanisms to repeatably deploy and test a distributed system for both integration and regression testing.

Testing ITBAs incurs additional difficulties that standard IoT applications do not face. For example, there is a completely separate network for each component of the actor in an ITBA: the IoT middleware/application layer, possibly a distributed database layer, and the blockchain layer.  We focus on the IoT middleware/application layer and blockchain layer in this paper. Testing thus requires that for each actor, we must run both the actor's IoT middleware/application code and a blockchain client instance. This pairing incurs a wide range of conditions that must be planned for, tracked, corrected, and tested. 

Some examples of what must be tested include: (1) the order of actor/blockchain client starting; (2) whether all actors should be started before processing on either the IoT and/or blockchain network starts; and (3) what detection and recovery mechanisms are needed to account for lost messages between the blockchain client and the actor, the actor losing a message, and transactions being lost on the blockchain. A testing environment for ITBAs, thus, needs to repeatably create networks and network conditions to address these conditions. Section~\ref{proposedArch} describes how the PlaTIBART architecture enables the building of such test networks.

\section{The Architecture of PlaTIBART}
\label{proposedArch}

PlaTIBART architecture for creating repeatable test network deployments of IoT/blockchain applications combines a Domain Specific Language (DSL) to define the network topology and settings, a Python program leveraging the Fabric API to manage the test network, and the RIAPS middleware~\cite{RIAPS_IEEE_ISORC_2017} to facilitate communication between nodes on the network. 
Each of these components is described below.

\subsection{Application Platform}


The \textit{Resilient Information Architecture Platform} for Smart Grid (RIAPS)~\cite{RIAPS_IEEE_ISORC_2017} is the application platform used by PlaTIBART to implement our case-study example described in Section~\ref{case_study}. 
RIAPS provides actor and component based abstraction, as well as support for deploying algorithms on devices across the network \footnote{RIAPS uses ZeroMQ \cite{hintjens2010zeromq} and Cap'n Proto \cite{varda2015cap} to manage the communication layer.} and solves problems collaboratively by providing micro-second level time synchronization \cite{RIAPS_IEEE_ISORC_2017}, failure based reconfiguration \cite{dubey2017resilience}, and group creation and coordination services (still under active development), in addition to the services described in \cite{LeeNiddodiSrivastavaBakken2016}. It is capable of handling different communications and running implemented algorithms in real-time.

\subsection{Actor Pattern}
\begin{figure}[ht]
\vspace{-0.10in}
	\centering
		
	\pgfdeclarelayer{background}
	\pgfsetlayers{background,main}

	\newcommand*\ab{.4}
	\tikzset{
		net node/.style = {circle, minimum width=2.75*\ab cm, inner sep=2pt, outer sep=0pt, draw},
		net node2/.style = {circle, minimum width=2.75*\ab cm, inner sep=7pt, outer sep=0pt, draw, fill=white!100},
		net root node/.style = {net node, minimum width=3*\ab cm},
		dashed connect/.style = {line width=2pt,dashed, draw=blue!50!cyan!25!black},
		net connect/.style = {line width=3pt, draw=blue!50!cyan!25!black},
		snake it/.style = {line width=2pt,decorate, decoration=snake, draw=blue!50!cyan!25!black},
		texter/.style = {draw, rectangle,anchor=west},
	}

	\begin{tikzpicture}
		
		\foreach \i in {0,...,4} {
			\path (-90+\i*72:2.25) node (i\i) [net node] {Wrapper};
			\path (-90+\i*72:4.5) node (o\i) [net node2] {geth};
			\path [net connect] (i\i) -- (o\i);
		
			\draw let \p1=(o\i), \p2=(i\i), \n1={atan2(\y2-\y1,\x2-\x1)}, \n2={veclen(\y2-\y1,\x2-\x1)}
			in ($ (o\i)!0.5!(i\i) $) ellipse [x radius=\n2/2+30pt, y radius=1.1cm,rotate=90-4*\n1];
		}
	
		\begin{pgfonlayer}{background}
			\foreach \x/\nodename in {o/snake it, i/dashed connect}
				\path [\nodename] (\x0) -- (\x1) -- (\x2) -- (\x3) -- (\x4) -- (\x0);
		\end{pgfonlayer}
	
		\newcommand{\y}{-3.2}
		\newcommand{\x}{-3.5}
		\newcommand{\yscale}{0.7}
		\newcommand{\keylinestart}{2.1}
		\newcommand{\keylineend}{1.5}
		
		\newcommand{\mylist} { dashed connect/.5/Off-Blockchain, net connect/2.5/Management APIs, snake it/1.5/On-Blockchain }
	
		\foreach \nodename/\spacer/\text in \mylist {
	
			\draw [\nodename] (\x-\keylineend,\y-\spacer*\yscale) -- (\x-\keylinestart,\y-\spacer*\yscale);	
			\draw [texter] (\x-1.2,\y-\spacer*\yscale) node {\text};
		}

		\end{tikzpicture}
	
    \vspace{-0.05in}
	\caption{Sample Actor Component Network with an Actor is a Geth Client and a Wrapper.}
	\label{fig:network-component-topology}
    \vspace{-0.10in}
\end{figure}
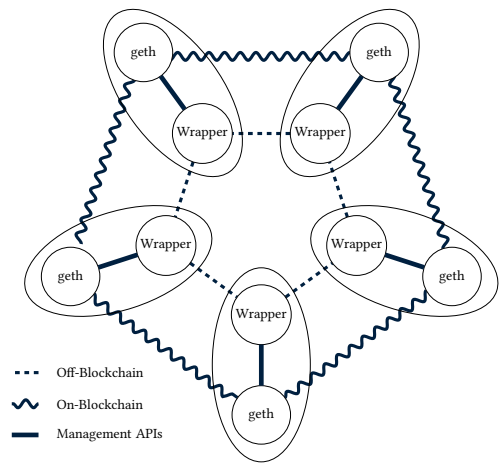

Each application client in the network is implemented as an actor with two main components: (1) a wrapper class specific to the role the actor is given and (2) a geth client, the  reference client for Ethereum\footnote{\url{https://github.com/ethereum/go-ethereum/wiki/geth}}. Figure \ref{fig:network-component-topology} shows a small network of five actors (indicated by an ellipse around a wrapper and geth client pair) and the networking connections between each actor's components. Geth clients communicate exclusively via on-blockchain means,\textit{ i.e.}, the geth client of each actor communicates directly with its associated wrapper, and the wrapper  communicates directly with other wrappers via an off-blockchain channel, such as TCP P2P communications. 

\subsection{Fault Tolerance}

A key benefit of decoupling the blockchain client and the wrapper into two  components of an actor is enhanced fault tolerance around transaction loss, compared with tightly coupled solutions. Specifically, it allows the wrapper to not only monitor the blockchain client, but also shut down and restart the client as needed. This design allows the wrapper component to ensure that if any known or discovered faults arise from defects in the blockchain software, the wrapper can at least attempt to recover. 

For example, in our Ethereum test network described in Section~\ref{experiment}, we have encountered faults where transactions are never mined~\cite{EthereumIssue14893:online} until a client is restarted. These lost transactions are problematic since they prevent a client from being able to interact with the blockchain network. Other types of faults, such as those related to an actor's communication with other components of the network, are handled by other middleware solutions, such as RIAPS.

PlaTIBART applies the \textit{Observer} pattern to notify the wrapper of the occurrence of events, such as faults and other blockchain-related conditions. This notification is accomplished by a separate thread within the wrapper that monitors its paired geth client for new events, such as completed transactions, or potential faults. This thread then notifies registered callback(s) when target events occur. For example, if the geth client becomes unresponsive or transactions appear to have stalled, then registered callback method(s) are called to notify the wrapper. 

\subsection{Domain Specific Language}

PlaTIBART's DSL defines the roles that different clients in our network have, based on the \textit{Actor} pattern. This DSL model implements a correct-by-construction design, thereby allowing for a verification stage on the model to check for internal consistency before any deployment is attempted. This verification prevents inconsistencies, such as two clients requesting the same port on the same host. 

Figure \ref{fig:model-sample} shows an example of our DSL, which specifies a full network configuration file for a test network. The first two lines of the configuration file contain two unique identifiers for this test network and its current version, ``configurationName'' and ``configurationVersion'', respectively. Next, it contains values specific for the creation of an Ethereum private network's Genesis block. 

A Genesis block in Ethereum is the first block in a blockchain and has special properties, such as not having a predecessor and being able to declare accounts that already have balances before any mining or transactions begin. The ``chainID'' is a unique positive integer identifying which blockchain the test network is using; 1 through 4 are public Ethereum blockchains of varying production/testing phases\footnote{By setting ``chainID'' to a public value, it is possible to connect a test network to a preexisting public blockchain network instead of creating a new one.}. 

Next, ``difficulty'' indicates how computationally hard it is to mine a block, and ``gasLimit'' is the maximum difficulty of a transaction based on length in bytes of the data and other Ethereum runtime values. The ``balance'' is the starting balance that we allocate to each client's starting account upon creation of the network\footnote{``balance'' applies only to accounts created before a new blockchain is created. Accounts created after the blockchain, be it public or private, is created will not receive any starting balance.}, which eliminates the situation where clients cannot begin transactions to request assets before any mining has begun. Lastly, the ``clients'' represent the actual nodes in our network. 

Clients in the DSL represent the individual actors in our network, comprised of a geth client and a RIAPs instance using a wrapper interface. The geth client has two interfaces and TCP ports associated with it: one for incoming Blockchain connections, and one for administration. 

\begin{figure}[t]
\begin{lstlisting}
{"configurationName":"test network a001",
"configurationVersion":"1",
"chainId": 15,
"difficulty": 100000,
"gasLimit": 200000000000000000,
"balance": 40000000000000000000000000,
"genesisBlockOutFile":"genesis-data.json",
"clients": {
    "startPort": 9000,
    "prosumer": {
        "count": 15,
        "hosts": [ "10.4.209.25",
                   "10.4.209.26",
                   "10.4.209.27",
                   "10.4.209.28" ] },
    "dso": {
        "count": 1,
        "hosts": [ "10.4.209.29" ] },
    "miner": {
        "count": 1,
        "hosts": [ "10.4.209.30" ] } } }
\end{lstlisting}
\vspace{-0.12in}
\caption{Sample DSL model.}
\label{fig:model-sample}
\vspace{-0.15in}
\end{figure}

\subsection{Network Manager}
\label{networkmanager}

The PlaTIBART network manager is written in Python and leverages the Fabric API to manage the SSH connections from a deployment/management host to each test host. The network manager is a stateless command-line program that takes a DSL file and a series of command operands to determine which operation it should perform. This design makes the manager readily adaptable into existing development workflows. 

Based on an input DSL file, PlaTIBART's network manager can create, start, stop, and delete ``clients'', ``miner''(s), and entire test network instances. The only exception to this approach is managing blockchains, which must have a configuration file made from the DSL file, and then adds that file into a new Genesis block. A Genesis block is the first block in a new Ethereum network. This Genesis block must be distributed to geth's private storage for each ``miner'' and ``clients'', before the test network can be started.


%

\section{Empirical Results}
\label{empirical}
This section examines our experimental testbed configuration and analyzes our experiment's results.
\subsection{Experimental Testbed Configuration}

To test PlaTIBART we implemented a solution to the Transactive Energy case study discussed in Section \ref{case_study} and deployed it to the test network defined in Figure \ref{fig:model-sample}. This network was installed on a private cloud instance hosted at Vanderbilt University. We ran our tests on 6 virtual hosts, each with: 4GB RAM, 40GB hard drive space, running Ubuntu 16.04.02, and gigabit networking.

For these tests we implemented a custom smart contract and wrappers for both Smart Grid distribution system operators (DSO) and prosumer clients in Python. Each wrapper had one geth client associated with it. We used PlaTIBART's network manager tool outlined in Section~\ref{networkmanager} to create, start, shutdown, and delete the test network. We manually paired each wrapper with its geth client's IP address and port (in future work this is to be integrated and automated into the network manager's capabilities).  

Using our custom written wrappers, smart contract, and managed test network we simulated a day's worth of transactive energy trading between actors. Via the Linux ``time'' command we measured each step needed in the entire process to create a test network, including Clients Create, Miners Create, Blockchain Make, Blockchain Create, Distribute to Clients, and Distribute to Miners. We also measured the steps required to start and connect the geth instance for each ``clients'' (``prosumer'' and ``DSO'') to the geth client of each ``miner.''\footnote{Miners are treated as a special case of ``clients'' and have their own unique set of network manager commands.} Currently, this star-network is the only network topology supported by PlaTIBART, but we will expand the supported topologies in the future. 

\subsection{Analysis of Results}
\label{results}
After running our tests, described above, we observed the results shown in Table~\ref{tab:1}. As shown in this table, the standard deviation for each testing phase was small (the largest being 0.09\% of the time taken). Likewise, the average time either remained relatively static, or scaled linearly, in relation to the number of clients (2, 5, 10, 15, 20 prosumers + 1 DSO) with one exception, that being ``network stop'', which had a sudden jump in its otherwise linear times at the 20 prosumer test.

\label{experiment}
\begin{table*}[!htbp]
	\centering
	\begin{tabular}{ l  r  r  r  r  r  r  r  r  r  r  } \hline

	&	\multicolumn{2}{c}{\textbf{2 Prosumers}}  	&	\multicolumn{2}{c}{\textbf{5 Prosumers}}	&	 \multicolumn{2}{c}{\textbf{10 Prosumers}}	&	\multicolumn{2}{c}{\textbf{15 Prosumers}} 	&	 \multicolumn{2}{c}{\textbf{20 Prosumers}}		\\ \hline

\textbf{Testing Stage}	&	\textbf{Avg} 	&	\textbf{Std Dev} 	&	\textbf{Avg} 	&	\textbf{Std Dev} 	&	\textbf{Avg} 	&	\textbf{Std Dev } 	&	\textbf{Avg} 	&	\textbf{Std Dev} 	&	\textbf{Avg} 	&	\textbf{Std Dev} 	\\ \hline

Clients Create	&	8.592 	&	0.172 	&	14.927 	&	0.097 	&	24.194 	&	0.160 	&	33.597 	&	0.194 	&	42.642 	&	0.285 	\\ \hline
Miners Create 	&	2.785 	&	0.011 	&	2.845 	&	0.066 	&	2.755 	&	0.008 	&	2.773 	&	0.042 	&	2.754 	&	0.029 	\\ \hline
Blockchain Make	&	0.069 	&	0.001 	&	0.070 	&	0.001 	&	0.071 	&	0.002 	&	0.072 	&	0.004 	&	0.070 	&	0.000 	\\ \hline
Blockchain Create	&	0.233 	&	0.010 	&	0.230 	&	0.020 	&	0.218 	&	0.018 	&	0.220 	&	0.009 	&	0.228 	&	0.015 	\\ \hline
Distribute to Clients	&	2.348 	&	0.012 	&	3.372 	&	0.064 	&	5.058 	&	0.063 	&	6.729 	&	0.111 	&	8.455 	&	0.061 	\\ \hline
Distribute to Miners	&	0.680 	&	0.017 	&	0.675 	&	0.009 	&	0.687 	&	0.016 	&	0.668 	&	0.021 	&	0.664 	&	0.023 	\\ \hline
Full Network Created	&	14.731 	&	0.157 	&	22.142 	&	0.188 	&	33.008 	&	0.217 	&	44.081 	&	0.137 	&	54.836 	&	0.337 	\\ \hline
Miner Start	&	0.800 	&	0.018 	&	0.806 	&	0.018 	&	0.801 	&	0.015 	&	0.805 	&	0.015 	&	0.811 	&	0.021 	\\ \hline
Clients Start	&	2.773 	&	0.019 	&	3.921 	&	0.085 	&	4.815 	&	0.099 	&	6.337 	&	0.132 	&	7.485 	&	0.258 	\\ \hline
Network Connect	&	0.504 	&	0.008 	&	0.932 	&	0.011 	&	1.634 	&	0.009 	&	2.401 	&	0.040 	&	3.071 	&	0.031 	\\ \hline
Network Stop	&	4.421 	&	0.034 	&	5.558 	&	0.034 	&	5.968 	&	0.085 	&	6.506 	&	0.088 	&	10.326 	&	0.249 	\\ \hline
Network Delete	&	5.332 	&	0.058 	&	5.288 	&	0.047 	&	5.290 	&	0.067 	&	5.297 	&	0.027 	&	5.446 	&	0.072 	\\ \hline
\end{tabular} 
	\caption{Average Time (Seconds) and Standard Deviation of Five Tests for Each Variation of Number of Prosumer Clients}
	\label{tab:1}
	\vspace{-0.3in}
\end{table*}

The test phases that remained relatively static included: Miners Create, Blockchain Make, Blockchain Create, Distribute to Miners, Miners Start, and Network Delete. The test phases that scaled with increase in number of prosumers were: Clients Create, Distribute to Clients, Full Network Created, Clients Start, Network Connect, and Network Stop. Other than Network Stop, which appears exponential but we suspect might become linear at greater network sizes, all of the scaling increases were linear (Std Dev < 0.065) after dividing the average time increase by the difference in number of clients. 

The results of our experiments indicate that there exists high consistency and predictability of managing PlaTIBART-based blockchain test networks, with the exception of the network stop operation, which needs further testing. These results help build confidence that PlaTIBART's approach to creating repeatable testing networks for IoT blockchain applications scales well, which is important to encourage adoption by IoT system developers. 


\section{Lessons Learned}
\label{lessonslearned}

During the implementation of our initial PlaTIBART prototype, we learned many lessons related to integrating IoT and blockchain. The three main categories of lessons learned involved documentation deficiencies, buggy behavior of the Ethereum geth client, and limitations of both Ethereum's management APIs and the Solidity smart contract language.

The official documentation for Ethereum is deficient in many key areas, such as organization, completeness, lack of meaningful examples, and clarity on best practices and security warnings. Here are some examples that demonstrate this:

\begin{itemize}
	\item Ethereum does not maintain its own documentation, instead linking to an outside resource maintained by volunteers from the Ethereum Community.
	\item The only official documentation is a FAQ on the main page, and the wikis in Ethereum's various source code repositories. 
	\item Important side effects of a management API call are only found listed under other method's documentation.
\end{itemize}

There are also programmatic bugs with Ethereum's reference client implementation, geth. While building and evaluating our test network, we experienced new transactions that were not mined regardless of how many new blocks were being mined, Ether available to the client, or any other obvious cause. 

Similar issues have been reported frequently on the public bug-reporting/tracking system about others attempting to setup private networks~\cite{EthereumIssue14893:online}. As of writing this paper, however, there is no solution other than to restart the geth client. This issue is addressable, but highlights the importance of fault tolerance in individual client execution and ways to recover from faults at that level. 

There are also idiosyncrasies of the Ethereum management APIs that are not well documented. An example is the polling mechanism that clients use to see if transactions occurred, which meet certain search criteria, and are called filters. The problem is that created filters are set to an undefined timer and will simply cease to work if not used ``for a while''~\cite{EthereumJSONRPC}. This quote, however, does not come from the description of the method for creating the filter. Instead, it is on a secondary method, eth\_uninstallFilter(), which is never referenced (directly or indirectly) from the original method, eth\_newFilter(). 

There are also limits to the Solidity smart contract language that must be accounted for in early planning stages of development. For example, the language currently does not support floating-point numbers. Moreover, all values must be converted to a specific binary representation for submission as a transaction. These limitations prevent---or dramatically increase the complexity (and therefore computational cost)---of advanced mathematical computations on-blockchain, which yields more off-blockchain processing and communication.

 
\section{Concluding Remarks}
\label{conclusion}
This paper describes how PlaTIBART applies the \textit{Actor} pattern with DSL-driven test network management software and component creation to enable the development of resilient, fault-tolerant IoT-blockchain applications and middleware. We employed PlaTIBART to dynamically deploy and manage test blockchain networks of varying sizes based on DSL configuration files. We also defined APIs for monitoring and recovering from faults, which standard blockchain applications were unable to recover from. This capability provides the means for fully integrated regression testing of blockchain applications, which is a novel contribution.

PlaTIBART currently  uses Ethereum as its blockchain implementation. For example, our DSL has Ethereum-specific required settings, such as ``chainId'' and ``gasLimit.'' Future versions of PlaTIBART will refactor these requirements so that other blockchain platforms, such as Hyperledger, can be substituted seamlessly.  Other areas of future work focus on formal verification of internal-consistency of a configuration file and a means of defining incremental adjustments to a test network through the DSL. Likewise, we are developing network management tools that help to simplify and automate the network topology for both the overall test framework instance, as well as which Actor components are paired.




\bibliographystyle{ACM-Reference-Format}
\bibliography{bibliography} 

\end{document}